\documentclass[twocolumn,superscriptaddress,showpacs,aps,amsmath,amssymb,prb]{revtex4}

\usepackage{bm}
\usepackage{graphicx,color}

\newcommand{\nl}{\nonumber \\}

\newcommand{\Sec}[1]{Sec.\;\ref{#1}}

\newcommand{\be}{\begin{equation}}
\newcommand{\ee}{\end{equation}}
\newcommand{\bea}{\begin{eqnarray}}
\newcommand{\eea}{\end{eqnarray}}
\newcommand{\bsube}{\begin{subequations}}
\newcommand{\esube}{\end{subequations}}
\newcommand{\Eq}[1]{Eq.\,(\ref{#1})}

\newcommand{\Fig}[1]{Fig.\,\ref{#1}}

\newcommand{\dg}{\dagger}
\newcommand{\la}{\langle}
\newcommand{\ra}{\rangle}

\begin{document}

\title{Electroluminescence and multi-photon effects in a resonator driven by 
 a tunnel junction}

\author{Jinshuang Jin}
\affiliation {Institute of Nanotechnology, Karlsruhe Institute of Technology (KIT), 76021 Karlsruhe, Germany}
\affiliation{ Department of Physics, Hangzhou Normal University,
  Hangzhou 310036, China}
\affiliation {Institut f\"ur Theoretische Festk\"orperphysik,
      Karlsruhe Institute of Technology (KIT), 76131 Karlsruhe, Germany}

\author{Michael Marthaler}
\affiliation {Institut f\"ur Theoretische Festk\"orperphysik,
      Karlsruhe Institute of Technology (KIT), 76131 Karlsruhe, Germany}

       \author{Gerd Sch\"on}
\affiliation {Institute of Nanotechnology, Karlsruhe Institute of Technology (KIT),  76021 Karlsruhe, Germany}
\affiliation {Institut f\"ur Theoretische Festk\"orperphysik,
      Karlsruhe Institute of Technology (KIT), 76131 Karlsruhe, Germany}
\affiliation{DFG Center for Functional Nanostructures, Karlsruhe Institute of Technology (KIT), 76131 Karlsruhe, Germany}

\date{\today}

\begin{abstract}
We consider a transmission line resonator which is driven by electrons tunneling through a 
voltage-biased tunnel junction.
Using the Born-Markovian quantum master equation in the polaron basis we investigate the nonequilibrium photon state and emission spectrum of the resonator as well as  properties of the transport current across the tunnel junction and its noise spectrum.
The electroluminescence is optimized, with maximum peak height and narrow linewidth,
when the back-action of the tunnel junction on the resonator and  the decay rate 
of the resonator are similar in strength. For strong coupling between the resonator and tunnel junction, multi-photon effects show up in the noise spectrum of the transport current.

\end{abstract}





\maketitle

\section{Introduction}
\label{intro}

Circuit quantum electrondynamics (cQED)
of on-chip solid-state systems coupled
to a microwave resonator has attracted 
much attention.
The investigations were stimulated by the possibility of strong coupling between a superconducting qubit 
   and a transmission line resonator \cite{Bla04062320,Wal04162,Chi04159}.
This allowed demonstrating phenomena known from quantum optics 
in the solid-state systems with unprecedented quality.
Some of the examples are vacuum Rabi splitting \cite{Wal04162,Chi04159},
further advanced applications of quantum state engineering~\cite{Maj07443,Sil07438,Hof09546}, as well as  
single-qubit lasing and cooling~\cite{Ast07588,Hau08037003,Gra08612}.

The development of cQED is not restricted to superconducting systems but has 
also been extended to solid-state devices composed of gate-defined semiconductor 
quantum dots or multi-dot systems coupled to
resonators~\cite{Chi04042302,Jin11035322,Fre11262105,Ber12045446,Fre12046807,Del11256804,Jin13025044}. 
Of particular interest is the interplay of the electrons tunneling through the dots and the 
excitation of photons in the resonator.
Single electron tunneling through a double
dot setup can produce a population inversion and induce a lasing state in the electromagnetic resonator
~\cite{Chi04042302,Jin11035322},
which is accompanied by pronounced features such as super- or sub-Poissonian noise of 
the transport current~\cite{Jin13025044}.
For a simpler system, a resonator driven by electrons tunneling through a single quantum dot,
the nonequilibrium photon population 
has also been investigated \cite{Ber12045446}.

Continuing to even more basic systems,  the question arises,
what is the nonequilibrium photon state created in the resonator by
electrons tunneling across a single junction without intermediate quantum dots.
Recently, such a system has been investigated experimentally,
and the resonator was found to influence the finite-frequency 
shot-noise of the transport current through the junction similar as a 
thermal electromagnetic environment~\cite{Alt14236803}.
A study of nonequilibrium effects in the resonator which is strongly 
coupled to a biased tunnel junction has,
to the best of our knowledge, not yet been performed, 
although the setup may find wide applications. For instance,
the system has been used as an effective charge detector 
for single-shot read-out of quantum-dot based qubits
 \cite{Gur9715215,Kor01115403,Elz04431,Fuj061634},
and as a displacement detector 
 which can resolve the momentum and position of nanomechanical resonators with high precision
\cite{Cle04121303, Pog08635,Ste1086,Wal11155411,Ben12075313}. 
Recently, it has been proposed that a tunnel junction can be used to generate squeezed light and microwave photon pairs~\cite{Gas13136601,For1403}.

  \begin{figure}[t]
 \centering
 \includegraphics[width=0.35\textwidth]{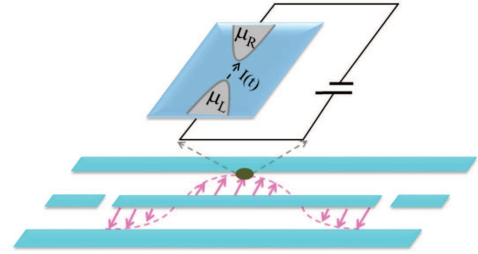} 
 \caption{(Color online) Schematic view of a tunnel junction--resonator circuit. 
The  junction is placed at a maximum of the electric field of the transmission line
resonator
  in order to maximize the dipole interaction.}
  \label{qpc-resonator}
 \end{figure}

In this work
we study the nonequilibrium photon state in a transmission line resonator which is strongly coupled to
the electrons tunneling through a tunnel junction (TJ). 
We focus on the electroluminescence of the excited photons
in the driven resonator,
 as well as the transport current through the tunnel junction and its noise spectrum.
In \Sec{thmeth},
we introduce the model of the TJ-resonator circuit
and present the quantum master equation 
describing the dynamics of the coupled system. 
We investigate the system in \Sec{thweakcoup} for 
 moderately strong coupling, where
single-photon processes dominate the dynamics. In this limit we find analytic results.
We then  study numerically in \Sec{thmulti} multi-photon effects which get visible 
in what is called the ultra-strong coupling limit. We conclude with a summary. 

\section{Methodology}
\label{thmeth}

\subsection{The system}
We consider a superconducting transmission line resonator strongly coupled to
a tunnel junction in a setup as sketched in Fig.~\ref{qpc-resonator}.
The corresponding Hamiltonian is given by ($\hbar=1$),
\begin{align}\label{H0}
H_{\rm tot}&=\sum_{\alpha k} \varepsilon_{\alpha k} c^\dg_{\alpha k} c_{\alpha k}
+\sum_{kk'}\big(t_{kk'}c^\dg_{Lk} c_{Rk'} +{\rm H.c.}\big)
\nl&\quad
+\omega_r a^\dg a+g \sum_k \big(c^\dg_{Rk} c_{Rk}- c^\dg_{Lk} c_{Lk} )(a+a^\dg).
\end{align}
The first line describes the tunnel junction between the left and right ($\alpha=L,R$) reservoirs with single-particle energies $\varepsilon_{\alpha k}$  
 and tunneling amplitudes $t_{kk'}$ between the two reservoirs.
 The resonator is modeled by a 
harmonic oscillator with frequency $\omega_r$.
The coupling of the two subsystems  with  strength $g$ 
is assumed to be induced by 
 the electric field of the resonator across the tunnel 
 junction, as illustrated in Fig.~\ref{qpc-resonator}, which
shifts the chemical potentials of the two reservoirs. 
We assume shifts of equal strength for both sides, but the generalization would be straightforward.

We proceed using the polaron transformation, $\widetilde H=U H U^\dg$ with
$U=\exp
\big[ \frac{g}{\omega_r} \sum_k( c^\dg_{Rk} c_{Rk}- c^\dg_{Lk} c_{Lk})(a^\dg-a)\big]$.
It transforms the Hamiltonian (\ref{H0}) to
\begin{align}\label{H1}
\widetilde H_{\rm tot}&=\sum_{\alpha k} \varepsilon_{\alpha k} c^\dg_{\alpha k} c_{\alpha k}
+\omega_r a^\dg a
\nl&\quad
+\sum_{kk'}
\big(t_{kk'}c^\dg_{Lk} c_{Rk'}  e^{-\lambda(a^\dg-a)}
+{\rm H.c.}\big) \, .
\end{align}
Here we neglected a trivial energy shift in the electrodes
and introduced the dimensionless coupling strength $\lambda\equiv2g/\omega_r$.
In the interaction picture with respect to the reservoir  
$H_{B}=\sum_{\alpha k}\varepsilon_{\alpha k} c^\dg_{\alpha k} c_{\alpha k}$, 
we recast the Hamiltonian (\ref{H1})  as $\widetilde H_{\rm tot}(t)=H_r+H'(t)$
with $H_r=\omega_r a^\dg a$ and coupling 
\begin{align}\label{Hcoup}
 H'(t)&=F^\dg(t) Q+Q^\dg F(t).
\end{align}
The operator of the tunnel junction is $F^\dg(t)=\sum_{kk'}t_{kk'}c^\dg_{Lk} c_{Rk'} e^{i\Delta_{kk'}t}$
with $\Delta_{kk'}=\varepsilon_{Lk}-\varepsilon_{Rk'}$, 
while $Q=\exp[-\lambda(a^\dg-a)]$ refers to the resonator. 
For later use we introduce
 the correlation functions of the bath accounting for forward (L to R) and backward tunneling,
 $C^{(+)}(t)\equiv \la F^{\dg}(t)F(0)\ra_{\rm B}$ and
$C^{(-)}(t)\equiv \la F(t)F^{\dg}(0)\ra_{\rm B}$, respectively. Here 
 $\la... \ra_{\rm B}$ stands for the statistical average over both
 electron reservoirs. They are assumed
to be in thermal equilibrium, in which case the correlators reduce to
\be\label{ct}
C^{(\pm)}(t) 
=\sum_{kk'}e^{\pm i\Delta_{kk'}t}|t_{kk'}|^2 f^{\pm}_{Lk}f^{\mp}_{Rk'}.
\ee
Here we introduced the Fermi-Dirac function of the $\alpha$-lead $f^+_{\alpha k}\equiv f_{\alpha k}=[e^{\beta(\varepsilon_{\alpha k}-\mu_\alpha)}+1]^{-1}$ 
with $\beta=1/(k_B T)$  and $f^{-}_{\alpha k}=1-f_{\alpha k}$. 
We focus on the limit of a tunnel junction, where the 
tunneling probabilities of each
channel are much smaller than unity, and we assume momentum-independent tunneling amplitudes
$t_{kk'}=t$. In combination with  the densities of states  $\nu_\alpha$  of the $\alpha$-reservoir
they determine the tunneling resistance $R$ and the dimensionless tunneling strength 
$\eta= 1/(2e^2 R) = \pi|t|^2 \nu_L\nu_R$. We assume $\eta \ll 1$ to be small.
The bath correlators in Fourier space
$\widetilde C^{(\pm)}(\omega)=\int^\infty_{-\infty} dt\, e^{i\omega t} 
  C^{(\pm)}(t)$ 
thus become  
 \begin{align}\label{cw}
  \widetilde C^{(\pm)}(\omega)=\frac{2\eta (\omega\pm eV)}{1-e^{-\beta(\omega\pm eV)}} \, .
 \end{align}
They account for forward and backward tunneling processes
with energy absorption ($\omega>0$) and emission ($\omega<0$).
Here $eV=\mu_L-\mu_R$ is the applied bias voltage across the tunnel junction.

\subsection{Quantum Master Equation}
Starting from  the total density operator $\rho_{\rm tot}(t)$ of the combined TJ-resonator system
one obtains the reduced density matrix of the resonator by tracing out the bath
degrees of freedom of the two electronic reservoirs,  $\rho(t)={\rm tr}_B[\rho_{\rm tot}(t)]$.
Treating $H'(t)$ as perturbation and expanding up to second-order leads to the Born-Markovian master equation  
\bsube\label{rhoeq}
\begin{align}
\label{rho0eq}
\dot\rho(t)
 &=-i[H_r,\rho(t)]+{\cal L}_{\kappa}\rho(t)+{\cal L}_{ B}\rho(t)\equiv{\cal L}\rho(t) \, .
\end{align}
While the first term describes the coherent evolution,
the second is the standard decay term of the resonator
with the decay rate $\kappa$, 
 and the third term accounts for the effect of the tunnel junction. 
They are given by 
\begin{align}\label{calLqpc0}
{\cal L}_{\kappa}\rho&=\kappa(n_{\rm th}+1)\big[a\rho a^\dg-\frac{1}{2}(a^\dg a \rho+\rho a^\dg a)\big]+
\nl&\quad
+\kappa n_{\rm th}\big[a^\dg\rho a-\frac{1}{2}(a a^\dg \rho+\rho a a^\dg)\big],
\\
{\cal L}_{B}\rho&=\frac{1}{2}\Big( \widetilde Q_-\rho Q^\dg+Q\rho \widetilde Q^\dg_-
-Q^\dg\widetilde Q_-\rho-\rho \widetilde Q^\dg_- Q
\nl&\quad\quad
+\widetilde Q^\dg_+\rho Q+Q^\dg\rho \widetilde Q_+
-Q\widetilde Q^\dg_+\rho-\rho \widetilde Q_+ Q^\dg
\Big)\, .
\end{align}
\esube
Here $n_{\rm th}=[\exp(\beta \omega_r)-1]^{-1}$ is the thermal photon number in the resonator,
and we introduced the operators 
\begin{align}
\widetilde Q_{\pm}&=
\int^\infty_{-\infty}\! dt\,C^{(\pm)}(t)e^{\pm i H_r t} Qe^{\mp i H_r t}. 
\nonumber
 \end{align}
The further calculations are done in the basis of Fock states, $H_r|n\ra=n\omega_r|n\ra$, of the photons in the resonator, for which the operator entering the coupling \Eq{Hcoup} is  expressed as
$Q=\sum_{nm}Q_{mn}|m\ra\la n|$ 
$Q_{mn}=\la m|e^{-\lambda(a^\dg-a)}|n\ra$. 
Correspondingly, the elements of the operator $\widetilde Q_{\pm}$ are calculated via
$\la m|\widetilde Q_{\pm}|n\ra=\widetilde  C^{(\pm)}(\pm \omega_{mn})Q_{mn}$,
with $\omega_{mn}\equiv(m-n)\omega_r$ and $\widetilde  C^{(\pm)}(\pm \omega_{mn})$ 
 given by \Eq{cw}.
%
Based on this master equation,
the emission spectrum of the resonator,
\be\label{Srw}
S_{\rm r}(\omega)\equiv \lim_{t\rightarrow\infty}\int^\infty_{-\infty} d\tau \la a^\dg(t) a(t+\tau) \ra e^{i\omega\tau}\, ,
\ee
as well as the second-order
correlation function in the stationary limit $g^{(2)}(\tau)=\lim_{t\rightarrow\infty}\la a^\dg(t)a^\dg(t+\tau)a(t+\tau)a(t)\ra/\la a^\dg(t)a(t)\ra^2$, 
can be calculated via the quantum regression theorem~\cite{Scu97}.

Starting from $I(t)= -e \, d\la n_R(t)\ra/dt$ with
$n_R=\sum_k c^\dg_{R k}c_{R k}$ we obtain
the transport current \cite{Li05205304}
$I(t)=\la \hat I(t)\ra ={\rm Tr}\big[ \hat I\rho(t)\big]$ with current operators 
\begin{align}\label{curr}
\hat{I}\rho(t)\!=\!\frac{e}{2}\left[
Q^\dg\rho(t)\widetilde Q_{+} -\widetilde Q_{-}\rho(t) Q^\dg+{\rm H.c.}\right].
\end{align}
From this we calculate the average current $I(t) $ and the 
 current noise spectrum  
 \be\label{curr_noise}
 S_I(\omega)= \int^\infty_{-\infty}
dt\, \la \{\delta \hat I(t),\delta \hat I(0)\}\ra e^{i\omega t},
\ee  
with $\delta \hat I(t)=\hat I(t)-I$. The noise spectrum can 
again be calculated using the quantum regression theorem.

 In the present work we consider a high-quality resonator with Q factor 
 assumed to be $2\times10^{4}$, corresponding to a 
   decay rate $\kappa=5\times10^{-5}\omega_r$.
   It is much smaller than both the tunneling rate and coupling strength, i.e., 
   $ \kappa/\omega_r\ll\eta,\lambda$.

\section{Moderate coupling strength}
\label{thweakcoup}


For weak to moderate coupling strength $\lambda \ll 1$  (but still $\kappa/\omega_r \ll \lambda$)
 we proceed in an  expansion up to 2nd order, i.e.
$Q=e^{-\lambda(a^\dg-a)}\approx 1-\lambda(a^\dg-a) +\frac{1}{2}\lambda^2(a^\dg-a)^2$. 
For definiteness we assume 
low temperatures, where the electrons tunnel only from the left
to the right lead without the reverse process, and the number of thermal
photons in the resonator vanishes, $n_{\rm th}=0$.
In this case the quantum master equation for the 
oscillator reduces to
\begin{align}
 \dot\rho\label{MELB}
 &=-i[H_r,\rho(t)]+
 (\kappa+\Gamma_{+})\big[a\rho a^\dg-\frac{1}{2}(a^\dg a \rho+\rho a^\dg a)\big]
 \nl&\quad
   +\Gamma_{-}\big[a^{\dag}\rho a-\frac{1}{2}(a a^{\dag} \rho+\rho a a^{\dag})\big],
\end{align}
with rates $\Gamma_{\pm}=\lambda^2\widetilde{C}^{(+)}(\pm\omega_r) $. 
In the considered limit we could make use of the rotating wave approximation.
The resulting master equation (\ref{MELB}) accounts for single-photon processes, i.e. processes where  electrons tunneling through the junction are associated with the
emission or absorption of a single photon in the resonator with rates $\Gamma_-$
and $\Gamma_+$, respectively. 
Interestingly, the second order term of the expansion of $Q$ does not modify the master equation, 
however, it does modify the average current to be studied later.

From \Eq{MELB} we see that the resonator is subject to an effective decay rate \cite{Car93}
\begin{align}\label{keff}
 {\kappa}_{\rm eff} &= \Gamma_+ -\Gamma_- +\kappa
           \approx 4\eta\lambda^2\omega_r+\kappa \,,
\end{align}
and the average photon number is
\begin{align}\label{aven}
 \bar n &=\frac{\Gamma_-}{\Gamma_+-\Gamma_-+\kappa}
           \approx\frac{\eta \lambda^2(eV-\omega_r)}{2\eta \lambda^2\omega_r+\kappa/2} \Theta(eV-\omega_r)\, .
\end{align}
Here, $\Theta$ is the step function, $\bar n=\la a^\dg a\ra=\sum_n nP_n$
with $P_n=\rho_{nn}$.
The corresponding  photon distribution,
$P_n\approx\la n\ra^n/(1+\la n\ra)^{n+1}$
coincides with a Bose-Einstein distribution with  effective temperature
\be\label{Teff}
k_BT_{\rm eff}\approx\omega_r/\ln\left[\frac{\eta\lambda^2(eV+\omega_r)+\kappa/2}
{\eta\lambda^2(eV-\omega_r)}\right] \Theta(eV-\omega_r).
\nonumber
\ee
This result coincides with the intensity distribution of classical
chaotic light \cite{Lou83,Bre02}.
A similar result has been obtained in Ref.\,\onlinecite{Ber12045446} for a resonator 
driven by electrons tunneling through a single quantum dot.

For the second-order correlation function we get 
\be
g^{(2)}(\tau)=1+e^{-\kappa_{\rm eff}\tau}.
\ee
It displays bunching, $g^{(2)}(0)=2$, for vanishing delay time and approaches $g^{(2)}(\tau\rightarrow\infty)=1$ for long delay time, 
when no correlations exist between the excited photons.

\begin{figure}
\centerline{\includegraphics*[width=1.0\columnwidth,angle=0]{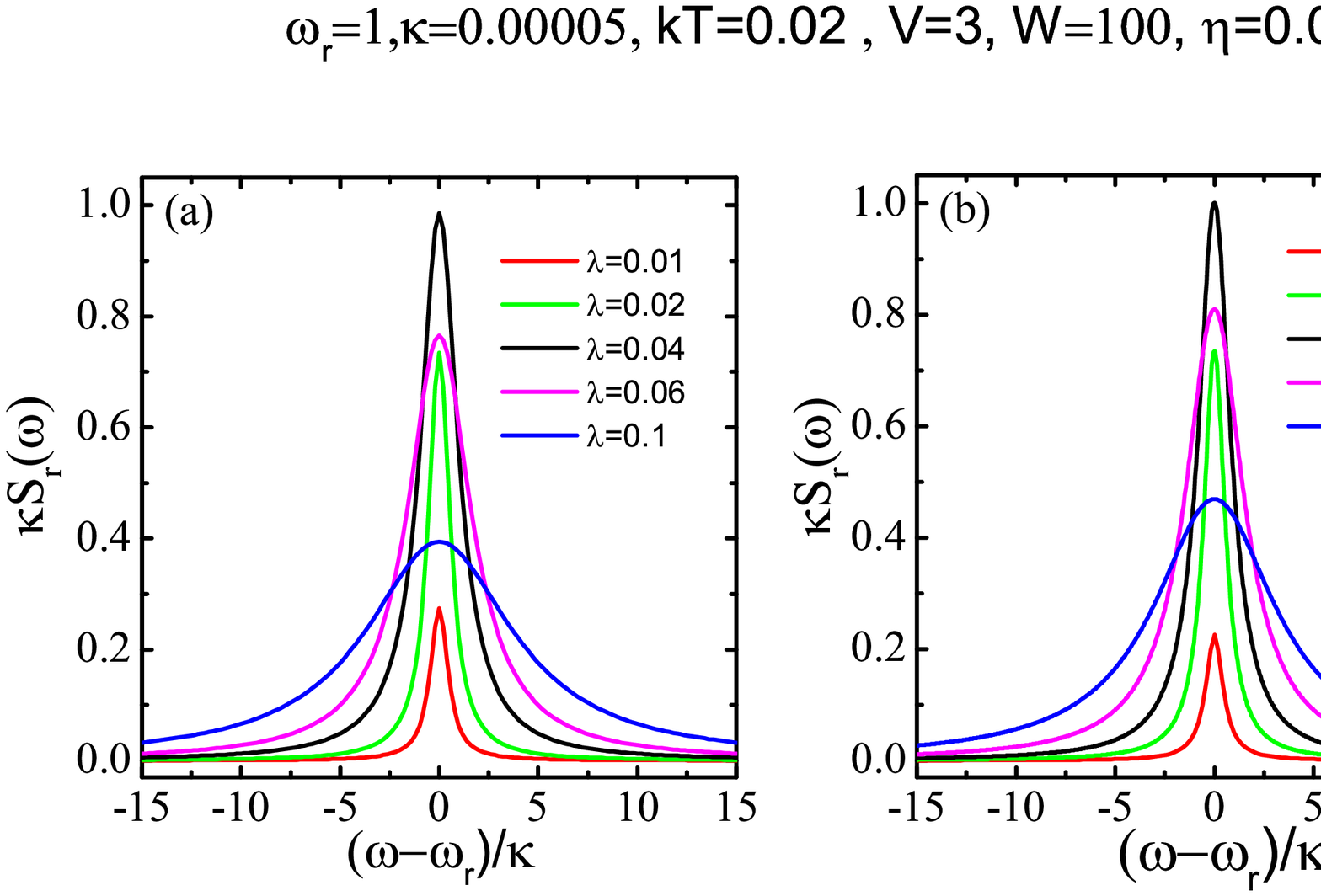}}
\caption{(Color online)
\label{fig_Srw}
The emission spectrum of the resonator $S_r(\omega)$ near the one-photon resonance, (a) for $\eta=0.01$ 
and different coupling strengths and (b) for $\lambda=0.02$ and
different tunneling rates. 
The other parameters are: low temperature $k_BT=0.02\omega_r$ and  bias voltage $eV=3\omega_r$.
 } 
\end{figure}

From the master equation we further get the emission spectrum of the resonator, 
\begin{equation}\label{Swr}
 S_{\rm r}(\omega)=\frac{\kappa_{\rm eff}\bar n}{(\omega_r-\omega)^2+(\kappa_{\rm eff}/2)^2}\, .
\end{equation}
Results are shown in 
\Fig{fig_Srw}.
With increasing coupling strength $\lambda$, or tunneling strength $\eta$, 
the height of the peak first increases  
 and then decreases with simultaneous broadening of the linewidth.
The maximum height of the peak  at $\omega=\omega_r$ is
\be\label{swr2}
  S_{\rm r}(\omega_r)= \frac{4\bar n}{\kappa_{\rm eff}}
  \approx\frac{2\eta\lambda^2(eV-\omega_r)}{(2\eta\lambda^2\omega_r+\kappa/2)^2} \Theta(eV-\omega_r).
  \ee
When the parameters satisfy the relation 
  \be\label{peak}
\eta_{\rm p}\lambda^2_{\rm p}= \frac{\kappa}{4\omega_r},
\ee  
the peak height is largest, with $S^{\rm max}_{\rm r}(\omega_r)=(eV-\omega_r)\Theta(eV-\omega_r)/(2\kappa\omega_r) $,
while the  linewidth is still narrow,
$(\kappa_{\rm eff})_{\rm p}/2=\kappa$. 
 This means that we find an optimal electroluminesence when
 the dissipative rate induced by the tunnel junction ($4\eta\lambda^2\omega_r$)
 is similar to the decay rate of the resonator ($\kappa$). 

\begin{figure}
\centerline{\includegraphics*[width=1.0\columnwidth,angle=0]{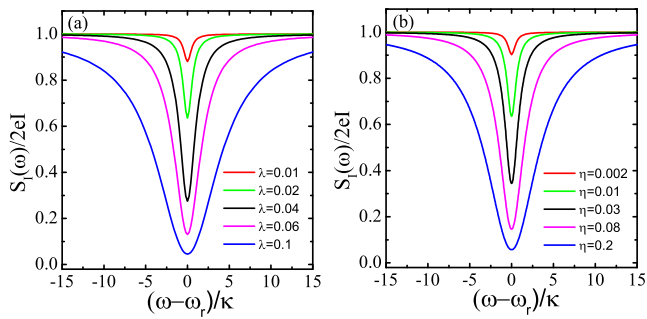}}
\caption{(Color online)
\label{fig_Sdw}
The noise spectrum of the transport current
through the junction $S_I(\omega)$, (a) for $\eta=0.01$ 
and different coupling strengths and (b) for $\lambda=0.02$ and
different tunneling rates. 
The other parameters are the same as in Fig.\,\ref{fig_Srw}.
} 
\end{figure}


 In the considered limit (i.e., up to $\lambda^2$) we get from \Eq{curr} and \Eq{curr_noise}
the average current, 
\begin{align}\label{avecurr}
 I& = (1-\lambda^2)\widetilde C^{(+)}(0) 
 +\lambda^2 \widetilde C^{(+)}(-\omega_r)  
 \nl&\approx
 2\eta (1-\lambda^2) eV+2\eta\lambda^2(eV-\omega_r) \Theta(eV-\omega_r),
\end{align}
 and
the current noise spectrum around $\omega=\pm\omega_r$, 
\be \label{Sw}
S_I(\omega)\!\approx\!2eI+\sum_{+,-}\frac{c_1\kappa_{\rm eff}/2}{(\omega\pm\omega_r)^2
+(\kappa_{\rm eff}/2)^2}\Theta(eV-\omega_r),
\ee
with the coefficient  
$c_1=-8e\eta^2\lambda^2 \omega_r[eV+(\bar n-1/2)\omega_r]$. 
Below the onset of single-photon processes the transport current is suppressed by the coupling to the resonator.
This effect is described by  
the Franck-Condon factor which renormalizes the tunneling rate \cite{Ber12045446} ,
 by a factor $(1-\frac{\lambda^2}{2})^2$.
Above the threshold, when photons can be exited the current grows as described by the
second term.
The noise spectrum further demonstrates the interplay of electrons tunneling
through the junction with the emission
and absorption of photons in the resonator.
This combination leads to a dip in the spectrum, shown in Fig.\,\ref{fig_Sdw}, at $\omega=\pm \omega_r$
. 
The dip gets deeper, i.e. $|S(\omega_r)-2eI|\approx 2|c_1|/\kappa_{\rm eff}$ 
increases, with growing coupling or tunneling strengths. 
The corresponding linewidth, $\kappa_{\rm eff}$
is consistent with that of the emission spectrum.

A comparison with the numerical solution of the full problem,
presented in the following section, shows that analytic results obtained so far for
weak to moderate coupling strength 
are valid as long as $\lambda\lesssim 0.2$.  

\begin{figure}
\centerline{\includegraphics*[width=1.02\columnwidth,angle=0]{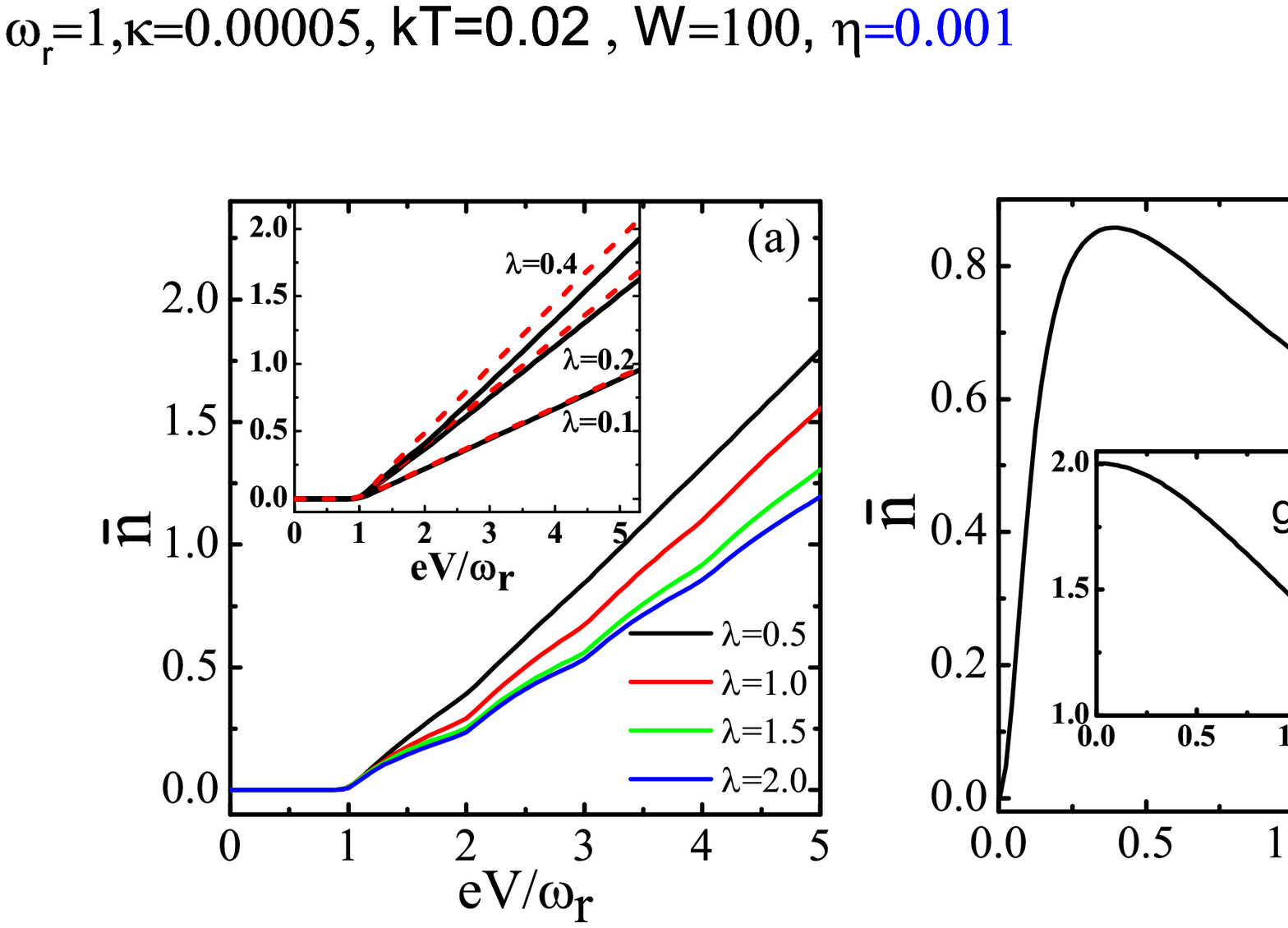}}
\caption{(Color online)
\label{fig_photon_str}
The average number of photons excited in the resonator 
(a) as a function of the bias voltage ($eV$)  and
(b) as a function of the coupling strength ($\lambda$) with $eV=3\omega_r$ 
at low temperature $k_BT=0.02\omega_r$ and tunneling rate $\eta=0.001$.
The insets in (a) is for moderate coupling strength
 based on the exact numerical calculation (solid-line)
and analytical expression of \Eq{aven} (dashed-line).
It is sufficient only for $\lambda \lesssim 0.2$. 
.
} 
\end{figure}

\section{Ultrastrong coupling}
\label{thmulti}
We turn now to the so-called ultrastrong-coupling regime where the coupling 
strength between  tunnel junction and resonator is of the order of the resonator
  frequency. 
 Values which we consider realistic,  and for which the present method is valid, are $0.1\omega_r<g\lesssim \omega_r$ (i.e., $0.2<\lambda=2g/\omega_r\lesssim 2$).
 Although more difficult to realize in an experiment, this limit displays interesting new properties.
%

In this regime, the single-photon approximation obtained from an expansion up to order $\lambda^2$ and analyzed in \Sec{thweakcoup}, is no longer sufficient. Instead two- and multi-photon processes associated with the excitations of multiple photons which follow from expanding
  $Q=e^{-\lambda(a^\dg-a)}$
to higher orders in $\lambda$ get important.
In order to study these processes we solved the equations introduced above numerically without further approximations.  
 In contrast to the single-photon limit, 
 the average photon number, as shown in Fig.\,\ref{fig_photon_str},
in general depends nonlinearly on the bias voltage 
  and even decreases with increasing coupling strength.
%
The nonequilibrium photon state is similar to that found when the resonator is 
driven by electrons tunneling through a single quantum dot, studied in Ref.\,\onlinecite{Ber12045446}. 
  The state of the photons in the resonator
deviates from a thermal state. E.g.,  as shown in the inset of Fig.\,\ref{fig_photon_str} (b) the second-order correlation function deviates from the value $g^{(2)}_{\rm thermal}(0)=2$ which we would find for a thermal (chaotic) state.

\begin{figure}
\centerline{\includegraphics*[width=1.0\columnwidth,angle=0]{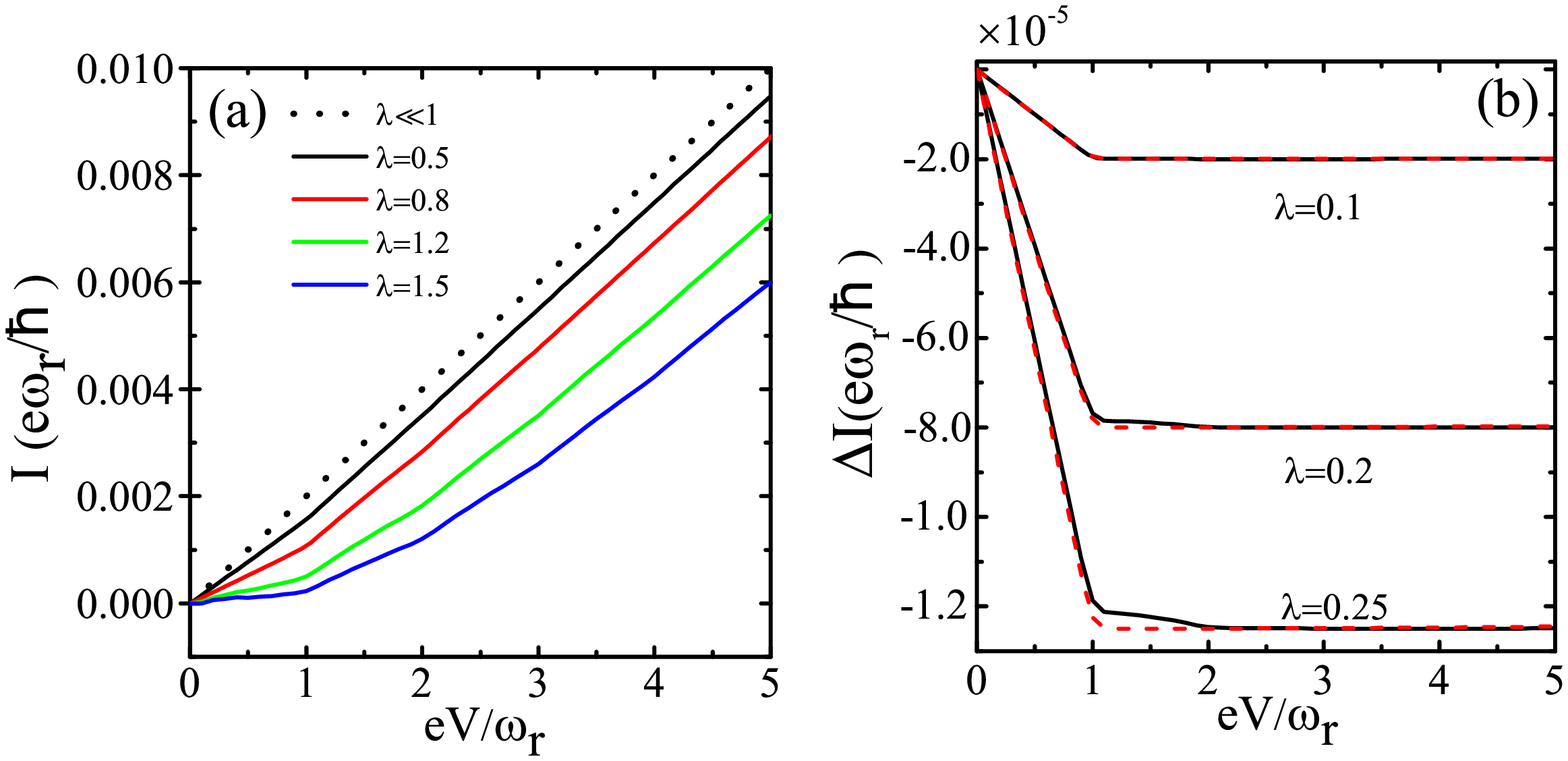}}
\caption{(Color online)
\label{fig_curr_str}
The average current tunneling through the junction as a function of the bias voltage
for different coupling strength. 
Multi-photon effect becomes significant with increasing  coupling strength shown in (a).
(b) displays the effect of the resonator on the transport current,
i.e., $\Delta I=I-I_0$ with $I_0=2\eta eV$, for moderate coupling strength
 based on the exact formula of \Eq{curr} (solid-line)
and analytical expression of \Eq{avecurr} (dashed-line).
The other parameters are the same as in Fig.\,\ref{fig_photon_str}}.
\end{figure}

\begin{figure}
\centerline{\includegraphics*[width=1.0\columnwidth,angle=0]{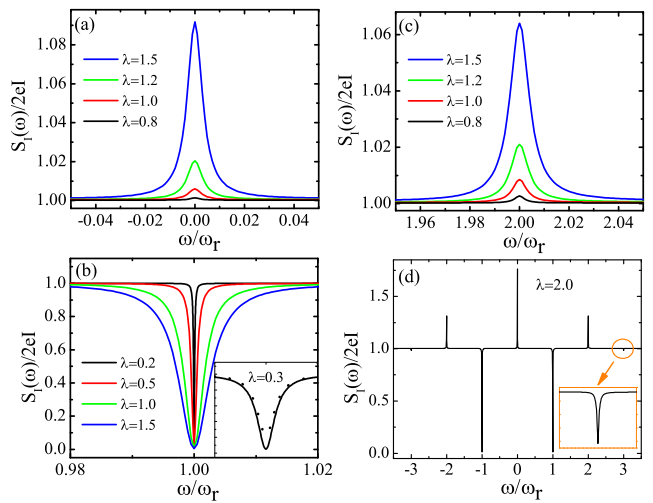}}
\caption{(Color online)
\label{fig_SIw_str}
 The noise spectrum of the transport current through the tunnel junction for ultrastrong-coupling 
 (a) around zero-frequency, (b) for single-photon, (c) two-photon, and (d) generally many-photon processes, respectively.
The dotted line in the insets of (c) is obtained in the weak/moderate coupling approximation \Eq{Sw}.
It well describes the single-photon process in the noise spectrum showing the dip behavior 
at $\omega=\omega_r$ for $\lambda\lesssim0.2$. 
The other parameters are the same as in Fig.\,\ref{fig_photon_str}.
 } 
\end{figure}

The effect of the multi-photon processes on the transport current becomes significant with increasing 
coupling strength and manifests itself in a nonlinear-dependence on the bias voltage
  as shown in Fig.\,\ref{fig_curr_str} (a).
Simultaneously, the multi-photon effects enhance the current fluctuations and
 induce the super-Poissonian behavior
in the zero-frequency shot noise
shown in Fig.\,\ref{fig_SIw_str} (a).
The multi-photon effects can also be observed in the current noise spectrum.
In an expansion up to fourth order in the coupling  
we obtain the noise spectrum near $\omega =0$  and $\omega = \pm 2 \omega_r$,
\begin{align}\label{Sw}
S_I(\omega)&\propto  e \eta^2\lambda^4\left[
\frac{c_0\kappa_{\rm eff}}{\omega^2+\kappa^2_{\rm eff}}
+\sum_{+,-}\frac{c_2\kappa_{\rm eff}}{(\omega\pm2\omega_r)^2+\kappa^2_{\rm eff}}\right],
\nonumber
\end{align}
with  positive coefficients 
$c_0>0$ 
 and $c_2>0$. 
The two-photon processes lead to peaks in the noise spectrum 
 at $\omega=0$ and $\omega=\pm 2\omega_r$ with linewidth determined by $\kappa_{\rm eff}$, as
shown in Fig.\,\ref{fig_SIw_str} (a) and (c).
Compared to the dip at $\omega=\omega_r$, the peaks at $\omega=0$ and 
 $\omega=\pm 2\omega_r$ are more sensitive to the coupling strength,
as the comparison of Figs.\,\ref{fig_SIw_str}(a), (b), and (c) demonstrates.
Three-photon effects, which we find by expanding further, 
lead again to a dip in the noise spectrum at $\omega=\pm 3\omega_r$, as
shown in Fig.\,\ref{fig_SIw_str} (d), with properties similar to the one-photon signal. 
We expect that the noise spectrum 
shows alternating dips and peaks for odd- (at $\omega=(2n+1)\omega_r$) and even-photon-number processes
(at $\omega=2n\omega_r$, $n=0,1,2\cdots$), respectively. 

\section{Summary}
\label{thsum}

In summary, we have  investigated the hybrid system 
of a transmission line resonator strongly coupled to a tunnel junction.
The study is based on a Born-Markov master equation in the polaron limit, which accounts  
for the nonequilibrium state of the resonator.
We presented results for two regimes of coupling strength between resonator and tunnel junction, 
characterized by single photon- or multiple photon-processes, respectively. 

For weak to moderate coupling, i.e., in the single-photon limit,
we obtained analytical results at low temperatures for
both the average number of the excited photons and the average current 
with threshold behavior once the bias voltage allows the excitation of photons.
The photon distribution can be parametrized by a thermal one with an enhanced effective temperature.
For the electroluminesence of the resonator we 
found the optimal conditions, with maximum height of the peak and still narrow linewidth,
when the resonator damping due to the tunnel junction is comparable in strength to
the intrinsic decay rate of the resonator. 
The current noise spectrum shows a pronounced dip at the resonator frequency.
These phenomena could be easily tested, since all the parameters are within
 reach of current experiments
 \cite{Wal04162,Fre11262105,Fre12046807,Alt14236803,Gas13136601,For1403,Ubb121}.

In the ultrastrong-coupling regime, 
multi-photon effects can be observed. 
The effect of the tunnel junction on the resonator can no longer be described by an effective heating.
The average number of photons excited in the
resonator, which first increases, eventually even decreases with increasing coupling strength.
The  multi-photon effects are most pronounced in the noise spectrum of the
transport current in the junction. In addition to the dips at $\omega=\pm \omega_r$
it shows peaks and dips 
at $\omega=\pm 2\omega_r$ and $\omega=\pm 3\omega_r$ and so forth due to
the interplay  of the electrons tunneling through the junction
associated with the emission and absorption of two-photons and three-photons 
 in the resonator, respectively.  
The current voltage characteristic shows threshold behavior at voltages $eV$ taking values which are multiples of the resonator frequency. 
While least spectacular this effect might be most easily observed in experiments.
 

\acknowledgments
We acknowledge stimulating discussions with K. Ensslin, T. Ihn, A. Wallraff, X.Q. Li, Y.J. Yan, P.-Q. Jin, D. Golubev and Andreas Heimes.
JSJ acknowledges support by a fellowship of the KIT, as well as the support in the 
Program of HNUEYT, and the NNSF of China (No.11274085).


\begin{thebibliography}{35}
\expandafter\ifx\csname natexlab\endcsname\relax\def\natexlab#1{#1}\fi
\expandafter\ifx\csname bibnamefont\endcsname\relax
  \def\bibnamefont#1{#1}\fi
\expandafter\ifx\csname bibfnamefont\endcsname\relax
  \def\bibfnamefont#1{#1}\fi
\expandafter\ifx\csname citenamefont\endcsname\relax
  \def\citenamefont#1{#1}\fi
\expandafter\ifx\csname url\endcsname\relax
  \def\url#1{\texttt{#1}}\fi
\expandafter\ifx\csname urlprefix\endcsname\relax\def\urlprefix{URL }\fi
\providecommand{\bibinfo}[2]{#2}
\providecommand{\eprint}[2][]{\url{#2}}

\bibitem[{\citenamefont{Blais et~al.}(2004)\citenamefont{Blais, Huang,
  Wallraff, Girvin, and Schoelkopf}}]{Bla04062320}
\bibinfo{author}{\bibfnamefont{A.}~\bibnamefont{Blais}},
  \bibinfo{author}{\bibfnamefont{R.-S.} \bibnamefont{Huang}},
  \bibinfo{author}{\bibfnamefont{A.}~\bibnamefont{Wallraff}},
  \bibinfo{author}{\bibfnamefont{S.~M.} \bibnamefont{Girvin}},
  \bibnamefont{and} \bibinfo{author}{\bibfnamefont{R.~J.}
  \bibnamefont{Schoelkopf}}, \bibinfo{journal}{Phys. Rev. A}
  \textbf{\bibinfo{volume}{69}}, \bibinfo{pages}{062320}
  (\bibinfo{year}{2004}).

\bibitem[{\citenamefont{Wallraff et~al.}(2004)\citenamefont{Wallraff, Schuster,
  Blais, Frunzio, R.~S.~Huang, Kumar, Girvin, and Schoelkopf}}]{Wal04162}
\bibinfo{author}{\bibfnamefont{A.}~\bibnamefont{Wallraff}},
  \bibinfo{author}{\bibfnamefont{D.~I.} \bibnamefont{Schuster}},
  \bibinfo{author}{\bibfnamefont{A.}~\bibnamefont{Blais}},
  \bibinfo{author}{\bibfnamefont{L.}~\bibnamefont{Frunzio}},
  \bibinfo{author}{\bibfnamefont{J.~M.} \bibnamefont{R.~S.~Huang}},
  \bibinfo{author}{\bibfnamefont{S.}~\bibnamefont{Kumar}},
  \bibinfo{author}{\bibfnamefont{S.~M.} \bibnamefont{Girvin}},
  \bibnamefont{and} \bibinfo{author}{\bibfnamefont{R.~J.}
  \bibnamefont{Schoelkopf}}, \bibinfo{journal}{Nature}
  \textbf{\bibinfo{volume}{431}}, \bibinfo{pages}{162} (\bibinfo{year}{2004}).

\bibitem[{\citenamefont{Chiorescu et~al.}(2004)\citenamefont{Chiorescu, Bertet,
  Semba, Nakamura, Harmans, and Mooij}}]{Chi04159}
\bibinfo{author}{\bibfnamefont{I.}~\bibnamefont{Chiorescu}},
  \bibinfo{author}{\bibfnamefont{P.}~\bibnamefont{Bertet}},
  \bibinfo{author}{\bibfnamefont{K.}~\bibnamefont{Semba}},
  \bibinfo{author}{\bibfnamefont{Y.}~\bibnamefont{Nakamura}},
  \bibinfo{author}{\bibfnamefont{C.~J. P.~M.} \bibnamefont{Harmans}},
  \bibnamefont{and} \bibinfo{author}{\bibfnamefont{J.~E.} \bibnamefont{Mooij}},
  \bibinfo{journal}{Nature} \textbf{\bibinfo{volume}{431}},
  \bibinfo{pages}{159} (\bibinfo{year}{2004}).

\bibitem[{\citenamefont{Majer et~al.}(2007)\citenamefont{Majer, Chow, Gambetta,
  Koch, Johnson, Schreier, Frunzio, Schuster, Houck, Wallraff
  et~al.}}]{Maj07443}
\bibinfo{author}{\bibfnamefont{J.}~\bibnamefont{Majer}},
  \bibinfo{author}{\bibfnamefont{J.~M.} \bibnamefont{Chow}},
  \bibinfo{author}{\bibfnamefont{J.~M.} \bibnamefont{Gambetta}},
  \bibinfo{author}{\bibfnamefont{J.}~\bibnamefont{Koch}},
  \bibinfo{author}{\bibfnamefont{B.~R.} \bibnamefont{Johnson}},
  \bibinfo{author}{\bibfnamefont{J.~A.} \bibnamefont{Schreier}},
  \bibinfo{author}{\bibfnamefont{L.}~\bibnamefont{Frunzio}},
  \bibinfo{author}{\bibfnamefont{D.~I.} \bibnamefont{Schuster}},
  \bibinfo{author}{\bibfnamefont{A.~A.} \bibnamefont{Houck}},
  \bibinfo{author}{\bibfnamefont{A.}~\bibnamefont{Wallraff}},
  \bibnamefont{et~al.}, \bibinfo{journal}{Nature}
  \textbf{\bibinfo{volume}{449}}, \bibinfo{pages}{443} (\bibinfo{year}{2007}).

\bibitem[{\citenamefont{Sillanp\"a\"a et~al.}(2007)\citenamefont{Sillanp\"a\"a,
  Park, and Simmonds}}]{Sil07438}
\bibinfo{author}{\bibfnamefont{M.~A.} \bibnamefont{Sillanp\"a\"a}},
  \bibinfo{author}{\bibfnamefont{J.~I.} \bibnamefont{Park}}, \bibnamefont{and}
  \bibinfo{author}{\bibfnamefont{R.~W.} \bibnamefont{Simmonds}},
  \bibinfo{journal}{Nature} \textbf{\bibinfo{volume}{449}},
  \bibinfo{pages}{438} (\bibinfo{year}{2007}).

\bibitem[{\citenamefont{Hofheinz et~al.}(2009)\citenamefont{Hofheinz, Wang,
  Ansmann, Bialczak, Lucero, Neeley, O'Connell, Sank, Wenner, Martinis
  et~al.}}]{Hof09546}
\bibinfo{author}{\bibfnamefont{M.}~\bibnamefont{Hofheinz}},
  \bibinfo{author}{\bibfnamefont{H.}~\bibnamefont{Wang}},
  \bibinfo{author}{\bibfnamefont{M.}~\bibnamefont{Ansmann}},
  \bibinfo{author}{\bibfnamefont{R.~C.} \bibnamefont{Bialczak}},
  \bibinfo{author}{\bibfnamefont{E.}~\bibnamefont{Lucero}},
  \bibinfo{author}{\bibfnamefont{M.}~\bibnamefont{Neeley}},
  \bibinfo{author}{\bibfnamefont{A.~D.} \bibnamefont{O'Connell}},
  \bibinfo{author}{\bibfnamefont{D.}~\bibnamefont{Sank}},
  \bibinfo{author}{\bibfnamefont{J.}~\bibnamefont{Wenner}},
  \bibinfo{author}{\bibfnamefont{J.~M.} \bibnamefont{Martinis}},
  \bibnamefont{et~al.}, \bibinfo{journal}{Nature}
  \textbf{\bibinfo{volume}{459}}, \bibinfo{pages}{546} (\bibinfo{year}{2009}).

\bibitem[{\citenamefont{Astafiev et~al.}(2007)\citenamefont{Astafiev, Inomata,
  Niskanen, Yamamoto, Pashkin, Nakamura, and Tsai}}]{Ast07588}
\bibinfo{author}{\bibfnamefont{O.}~\bibnamefont{Astafiev}},
  \bibinfo{author}{\bibfnamefont{K.}~\bibnamefont{Inomata}},
  \bibinfo{author}{\bibfnamefont{A.~O.} \bibnamefont{Niskanen}},
  \bibinfo{author}{\bibfnamefont{T.}~\bibnamefont{Yamamoto}},
  \bibinfo{author}{\bibfnamefont{Y.~A.} \bibnamefont{Pashkin}},
  \bibinfo{author}{\bibfnamefont{Y.}~\bibnamefont{Nakamura}}, \bibnamefont{and}
  \bibinfo{author}{\bibfnamefont{J.~S.} \bibnamefont{Tsai}},
  \bibinfo{journal}{Nature} \textbf{\bibinfo{volume}{449}},
  \bibinfo{pages}{588} (\bibinfo{year}{2007}).

\bibitem[{\citenamefont{Hauss et~al.}(2008)\citenamefont{Hauss, Fedorov,
  Hutter, Shnirman, and Sch\"on}}]{Hau08037003}
\bibinfo{author}{\bibfnamefont{J.}~\bibnamefont{Hauss}},
  \bibinfo{author}{\bibfnamefont{A.}~\bibnamefont{Fedorov}},
  \bibinfo{author}{\bibfnamefont{C.}~\bibnamefont{Hutter}},
  \bibinfo{author}{\bibfnamefont{A.}~\bibnamefont{Shnirman}}, \bibnamefont{and}
  \bibinfo{author}{\bibfnamefont{G.}~\bibnamefont{Sch\"on}},
  \bibinfo{journal}{Phys. Rev. Lett.} \textbf{\bibinfo{volume}{100}},
  \bibinfo{pages}{037003} (\bibinfo{year}{2008}).

\bibitem[{\citenamefont{Grajcar et~al.}(2008)\citenamefont{Grajcar, van~der
  Ploeg, Izmalkov, E.~Ilichev, Fedorov, Shnirman, and {Sch\"{o}n}}}]{Gra08612}
\bibinfo{author}{\bibfnamefont{M.}~\bibnamefont{Grajcar}},
  \bibinfo{author}{\bibfnamefont{S.~H.~W.} \bibnamefont{van~der Ploeg}},
  \bibinfo{author}{\bibfnamefont{A.}~\bibnamefont{Izmalkov}},
  \bibinfo{author}{\bibfnamefont{H.~G.~M.} \bibnamefont{E.~Ilichev}},
  \bibinfo{author}{\bibfnamefont{A.}~\bibnamefont{Fedorov}},
  \bibinfo{author}{\bibfnamefont{A.}~\bibnamefont{Shnirman}}, \bibnamefont{and}
  \bibinfo{author}{\bibfnamefont{G.}~\bibnamefont{{Sch\"{o}n}}},
  \bibinfo{journal}{Nature Physics} \textbf{\bibinfo{volume}{4}},
  \bibinfo{pages}{612} (\bibinfo{year}{2008}).

\bibitem[{\citenamefont{Childress et~al.}(2004)\citenamefont{Childress,
  S\o{}rensen, and Lukin}}]{Chi04042302}
\bibinfo{author}{\bibfnamefont{L.}~\bibnamefont{Childress}},
  \bibinfo{author}{\bibfnamefont{A.~S.} \bibnamefont{S\o{}rensen}},
  \bibnamefont{and} \bibinfo{author}{\bibfnamefont{M.~D.} \bibnamefont{Lukin}},
  \bibinfo{journal}{Phys. Rev. A} \textbf{\bibinfo{volume}{69}},
  \bibinfo{pages}{042302} (\bibinfo{year}{2004}).

\bibitem[{\citenamefont{Jin et~al.}(2011)\citenamefont{Jin, Marthaler, Cole,
  Shnirman, and Sch\"on}}]{Jin11035322}
\bibinfo{author}{\bibfnamefont{P.-Q.} \bibnamefont{Jin}},
  \bibinfo{author}{\bibfnamefont{M.}~\bibnamefont{Marthaler}},
  \bibinfo{author}{\bibfnamefont{J.~H.} \bibnamefont{Cole}},
  \bibinfo{author}{\bibfnamefont{A.}~\bibnamefont{Shnirman}}, \bibnamefont{and}
  \bibinfo{author}{\bibfnamefont{G.}~\bibnamefont{Sch\"on}},
  \bibinfo{journal}{Phys. Rev. B} \textbf{\bibinfo{volume}{84}},
  \bibinfo{pages}{035322} (\bibinfo{year}{2011}).


\bibitem[{\citenamefont{Frey et~al.}(2011)\citenamefont{Frey, Leek, Beck,
  Ensslin, Wallraff, and Ihn}}]{Fre11262105}
\bibinfo{author}{\bibfnamefont{T.}~\bibnamefont{Frey}},
  \bibinfo{author}{\bibfnamefont{P.~J.} \bibnamefont{Leek}},
  \bibinfo{author}{\bibfnamefont{M.}~\bibnamefont{Beck}},
  \bibinfo{author}{\bibfnamefont{K.}~\bibnamefont{Ensslin}},
  \bibinfo{author}{\bibfnamefont{A.}~\bibnamefont{Wallraff}}, \bibnamefont{and}
  \bibinfo{author}{\bibfnamefont{T.}~\bibnamefont{Ihn}},
  \bibinfo{journal}{Applied Physics Letters} \textbf{\bibinfo{volume}{98}},
  \bibinfo{eid}{262105} (\bibinfo{year}{2011}).
  

\bibitem[{\citenamefont{Bergenfeldt and Samuelsson}(2012)}]{Ber12045446}
\bibinfo{author}{\bibfnamefont{C.}~\bibnamefont{Bergenfeldt}} \bibnamefont{and}
  \bibinfo{author}{\bibfnamefont{P.}~\bibnamefont{Samuelsson}},
  \bibinfo{journal}{Phys. Rev. B} \textbf{\bibinfo{volume}{85}},
  \bibinfo{pages}{045446} (\bibinfo{year}{2012}).



\bibitem[{\citenamefont{Frey et~al.}(2012)\citenamefont{Frey, Leek, Beck,
  Blais, Ihn, Ensslin, and Wallraff}}]{Fre12046807}
\bibinfo{author}{\bibfnamefont{T.}~\bibnamefont{Frey}},
  \bibinfo{author}{\bibfnamefont{P.~J.} \bibnamefont{Leek}},
  \bibinfo{author}{\bibfnamefont{M.}~\bibnamefont{Beck}},
  \bibinfo{author}{\bibfnamefont{A.}~\bibnamefont{Blais}},
  \bibinfo{author}{\bibfnamefont{T.}~\bibnamefont{Ihn}},
  \bibinfo{author}{\bibfnamefont{K.}~\bibnamefont{Ensslin}}, \bibnamefont{and}
  \bibinfo{author}{\bibfnamefont{A.}~\bibnamefont{Wallraff}},
  \bibinfo{journal}{Phys. Rev. Lett.} \textbf{\bibinfo{volume}{108}},
  \bibinfo{pages}{046807} (\bibinfo{year}{2012}).

\bibitem[{\citenamefont{Delbecq et~al.}(2011)\citenamefont{Delbecq, Schmitt,
  Parmentier, Roch, Viennot, F\`eve, Huard, Mora, Cottet, and
  Kontos}}]{Del11256804}
\bibinfo{author}{\bibfnamefont{M.~R.} \bibnamefont{Delbecq}},
  \bibinfo{author}{\bibfnamefont{V.}~\bibnamefont{Schmitt}},
  \bibinfo{author}{\bibfnamefont{F.~D.} \bibnamefont{Parmentier}},
  \bibinfo{author}{\bibfnamefont{N.}~\bibnamefont{Roch}},
  \bibinfo{author}{\bibfnamefont{J.~J.} \bibnamefont{Viennot}},
  \bibinfo{author}{\bibfnamefont{G.}~\bibnamefont{F\`eve}},
  \bibinfo{author}{\bibfnamefont{B.}~\bibnamefont{Huard}},
  \bibinfo{author}{\bibfnamefont{C.}~\bibnamefont{Mora}},
  \bibinfo{author}{\bibfnamefont{A.}~\bibnamefont{Cottet}}, \bibnamefont{and}
  \bibinfo{author}{\bibfnamefont{T.}~\bibnamefont{Kontos}},
  \bibinfo{journal}{Phys. Rev. Lett.} \textbf{\bibinfo{volume}{107}},
  \bibinfo{pages}{256804} (\bibinfo{year}{2011}).

\bibitem[{\citenamefont{Jin et~al.}(2013)\citenamefont{Jin, Marthaler, Jin,
  Golubev, and Sch\"on}}]{Jin13025044}
\bibinfo{author}{\bibfnamefont{J.}~\bibnamefont{Jin}},
  \bibinfo{author}{\bibfnamefont{M.}~\bibnamefont{Marthaler}},
  \bibinfo{author}{\bibfnamefont{P.-Q.} \bibnamefont{Jin}},
  \bibinfo{author}{\bibfnamefont{D.}~\bibnamefont{Golubev}}, \bibnamefont{and}
  \bibinfo{author}{\bibfnamefont{G.}~\bibnamefont{Sch\"on}},
  \bibinfo{journal}{New Journal of Physics} \textbf{\bibinfo{volume}{15}},
  \bibinfo{pages}{025044} (\bibinfo{year}{2013}).

\bibitem[{\citenamefont{Altimiras
  et~al.}(2014{\natexlab{a}})\citenamefont{Altimiras, Parlavecchio, Joyez,
  Vion, Roche, Esteve, and Portier}}]{Alt14236803}
\bibinfo{author}{\bibfnamefont{C.}~\bibnamefont{Altimiras}},
  \bibinfo{author}{\bibfnamefont{O.}~\bibnamefont{Parlavecchio}},
  \bibinfo{author}{\bibfnamefont{P.}~\bibnamefont{Joyez}},
  \bibinfo{author}{\bibfnamefont{D.}~\bibnamefont{Vion}},
  \bibinfo{author}{\bibfnamefont{P.}~\bibnamefont{Roche}},
  \bibinfo{author}{\bibfnamefont{D.}~\bibnamefont{Esteve}}, \bibnamefont{and}
  \bibinfo{author}{\bibfnamefont{F.}~\bibnamefont{Portier}},
  \bibinfo{journal}{Phys. Rev. Lett.} \textbf{\bibinfo{volume}{112}},
  \bibinfo{pages}{236803} (\bibinfo{year}{2014}{\natexlab{a}}).

\bibitem[{\citenamefont{Gurvitz}(1997)}]{Gur9715215}
\bibinfo{author}{\bibfnamefont{S.~A.} \bibnamefont{Gurvitz}},
  \bibinfo{journal}{Phys. Rev. B} \textbf{\bibinfo{volume}{56}},
  \bibinfo{pages}{15215} (\bibinfo{year}{1997}).

\bibitem[{\citenamefont{Korotkov}(2001)}]{Kor01115403}
\bibinfo{author}{\bibfnamefont{A.~N.} \bibnamefont{Korotkov}},
  \bibinfo{journal}{Phys. Rev. B} \textbf{\bibinfo{volume}{63}},
  \bibinfo{pages}{115403} (\bibinfo{year}{2001}).

\bibitem[{\citenamefont{Elzerman et~al.}(2004)\citenamefont{Elzerman, Hanson,
  {Willems van Beveren}, Witkamp, Vandersypen, and Kouwenhoven}}]{Elz04431}
\bibinfo{author}{\bibfnamefont{J.~M.} \bibnamefont{Elzerman}},
  \bibinfo{author}{\bibfnamefont{R.}~\bibnamefont{Hanson}},
  \bibinfo{author}{\bibfnamefont{L.~H.} \bibnamefont{{Willems van Beveren}}},
  \bibinfo{author}{\bibfnamefont{B.}~\bibnamefont{Witkamp}},
  \bibinfo{author}{\bibfnamefont{L.~M.~K.} \bibnamefont{Vandersypen}},
  \bibnamefont{and} \bibinfo{author}{\bibfnamefont{L.~P.}
  \bibnamefont{Kouwenhoven}}, \bibinfo{journal}{Nature}
  \textbf{\bibinfo{volume}{430}}, \bibinfo{pages}{431} (\bibinfo{year}{2004}).

\bibitem[{\citenamefont{Fujisawa et~al.}(2006)\citenamefont{Fujisawa, Hayashi,
  Tomita, and Hirayama}}]{Fuj061634}
\bibinfo{author}{\bibfnamefont{T.}~\bibnamefont{Fujisawa}},
  \bibinfo{author}{\bibfnamefont{T.}~\bibnamefont{Hayashi}},
  \bibinfo{author}{\bibfnamefont{R.}~\bibnamefont{Tomita}}, \bibnamefont{and}
  \bibinfo{author}{\bibfnamefont{Y.}~\bibnamefont{Hirayama}},
  \textbf{\bibinfo{volume}{312}}, \bibinfo{pages}{1634} (\bibinfo{year}{2006}).

\bibitem[{\citenamefont{Clerk and Girvin}(2004)}]{Cle04121303}
\bibinfo{author}{\bibfnamefont{A.~A.} \bibnamefont{Clerk}} \bibnamefont{and}
  \bibinfo{author}{\bibfnamefont{S.~M.} \bibnamefont{Girvin}},
  \bibinfo{journal}{Phys. Rev. B} \textbf{\bibinfo{volume}{70}},
  \bibinfo{pages}{121303} (\bibinfo{year}{2004}).

\bibitem[{\citenamefont{Poggio et~al.}(2008)\citenamefont{Poggio, Jura, Degen,
  Topinka, Mamin, Goldhaber-Gordon, and Rugar}}]{Pog08635}
\bibinfo{author}{\bibfnamefont{M.}~\bibnamefont{Poggio}},
  \bibinfo{author}{\bibfnamefont{M.~P.} \bibnamefont{Jura}},
  \bibinfo{author}{\bibfnamefont{C.~L.} \bibnamefont{Degen}},
  \bibinfo{author}{\bibfnamefont{M.~A.} \bibnamefont{Topinka}},
  \bibinfo{author}{\bibfnamefont{H.~J.} \bibnamefont{Mamin}},
  \bibinfo{author}{\bibfnamefont{D.}~\bibnamefont{Goldhaber-Gordon}},
  \bibnamefont{and} \bibinfo{author}{\bibfnamefont{D.}~\bibnamefont{Rugar}},
  \bibinfo{journal}{Nat. Phys.} \textbf{\bibinfo{volume}{4}},
  \bibinfo{pages}{635} (\bibinfo{year}{2008}).

\bibitem[{\citenamefont{Stettenheim et~al.}(2010)\citenamefont{Stettenheim,
  Thalakulam, Pan, Bal, Ji, Xue, Pfeiffer, West, Blencowe, and
  Rimberg}}]{Ste1086}
\bibinfo{author}{\bibfnamefont{J.}~\bibnamefont{Stettenheim}},
  \bibinfo{author}{\bibfnamefont{M.}~\bibnamefont{Thalakulam}},
  \bibinfo{author}{\bibfnamefont{F.}~\bibnamefont{Pan}},
  \bibinfo{author}{\bibfnamefont{M.}~\bibnamefont{Bal}},
  \bibinfo{author}{\bibfnamefont{Z.}~\bibnamefont{Ji}},
  \bibinfo{author}{\bibfnamefont{W.}~\bibnamefont{Xue}},
  \bibinfo{author}{\bibfnamefont{L.}~\bibnamefont{Pfeiffer}},
  \bibinfo{author}{\bibfnamefont{K.~W.} \bibnamefont{West}},
  \bibinfo{author}{\bibfnamefont{M.~P.} \bibnamefont{Blencowe}},
  \bibnamefont{and} \bibinfo{author}{\bibfnamefont{A.~J.}
  \bibnamefont{Rimberg}}, \bibinfo{journal}{Nature Letters}
  \textbf{\bibinfo{volume}{466}}, \bibinfo{pages}{86} (\bibinfo{year}{2010}).

\bibitem[{\citenamefont{Walter and Trauzettel}(2011)}]{Wal11155411}
\bibinfo{author}{\bibfnamefont{S.}~\bibnamefont{Walter}} \bibnamefont{and}
  \bibinfo{author}{\bibfnamefont{B.}~\bibnamefont{Trauzettel}},
  \bibinfo{journal}{Phys. Rev. B} \textbf{\bibinfo{volume}{83}},
  \bibinfo{pages}{155411} (\bibinfo{year}{2011}).

\bibitem[{\citenamefont{Benatov and Blencowe}(2012)}]{Ben12075313}
\bibinfo{author}{\bibfnamefont{L.~L.} \bibnamefont{Benatov}} \bibnamefont{and}
  \bibinfo{author}{\bibfnamefont{M.~P.} \bibnamefont{Blencowe}},
  \bibinfo{journal}{Phys. Rev. B} \textbf{\bibinfo{volume}{86}},
  \bibinfo{pages}{075313} (\bibinfo{year}{2012}).

\bibitem[{\citenamefont{Gasse et~al.}(2013)\citenamefont{Gasse, Lupien, and
  Reulet}}]{Gas13136601}
\bibinfo{author}{\bibfnamefont{G.}~\bibnamefont{Gasse}},
  \bibinfo{author}{\bibfnamefont{C.}~\bibnamefont{Lupien}}, \bibnamefont{and}
  \bibinfo{author}{\bibfnamefont{B.}~\bibnamefont{Reulet}},
  \bibinfo{journal}{Phys. Rev. Lett.} \textbf{\bibinfo{volume}{111}},
  \bibinfo{pages}{136601} (\bibinfo{year}{2013}).

\bibitem[{\citenamefont{Forgues et~al.}(2014)\citenamefont{Forgues, Lupien, and
  Reulet}}]{For1403}
\bibinfo{author}{\bibfnamefont{J.-C.} \bibnamefont{Forgues}},
  \bibinfo{author}{\bibfnamefont{C.}~\bibnamefont{Lupien}}, \bibnamefont{and}
  \bibinfo{author}{\bibfnamefont{B.}~\bibnamefont{Reulet}},
  \bibinfo{journal}{eprint} p. \bibinfo{pages}{arXiv:1403.5578}
  (\bibinfo{year}{2014}).

\bibitem[{\citenamefont{Scully and Zubairy}(1997)}]{Scu97}
\bibinfo{author}{\bibfnamefont{M.}~\bibnamefont{Scully}} \bibnamefont{and}
  \bibinfo{author}{\bibfnamefont{M.~S.} \bibnamefont{Zubairy}},
  \emph{\bibinfo{title}{Quantum Optics}} (\bibinfo{publisher}{Cambridge
  University Press}, \bibinfo{address}{Cambridge}, \bibinfo{year}{1997}).

\bibitem[{\citenamefont{Li et~al.}(2005)\citenamefont{Li, Luo, Yang, Cui, and
  Yan}}]{Li05205304}
\bibinfo{author}{\bibfnamefont{X.~Q.} \bibnamefont{Li}},
  \bibinfo{author}{\bibfnamefont{J.~Y.} \bibnamefont{Luo}},
  \bibinfo{author}{\bibfnamefont{Y.~G.} \bibnamefont{Yang}},
  \bibinfo{author}{\bibfnamefont{P.}~\bibnamefont{Cui}}, \bibnamefont{and}
  \bibinfo{author}{\bibfnamefont{Y.~J.} \bibnamefont{Yan}},
  \bibinfo{journal}{Phys. Rev. B} \textbf{\bibinfo{volume}{71}},
  \bibinfo{pages}{205304} (\bibinfo{year}{2005}).

\bibitem[{\citenamefont{Carmichael}(1993)}]{Car93}
\bibinfo{author}{\bibfnamefont{H.~J.} \bibnamefont{Carmichael}},
  \emph{\bibinfo{title}{An Open System Approach to Quantum Optics}}
  (\bibinfo{publisher}{Spring-Verlag}, \bibinfo{address}{Berlin},
  \bibinfo{year}{1993}).

\bibitem[{\citenamefont{Loudon}(1983)}]{Lou83}
\bibinfo{author}{\bibfnamefont{R.}~\bibnamefont{Loudon}},
  \emph{\bibinfo{title}{The Quantum Theory of Light}}
  (\bibinfo{publisher}{Clarendon}, \bibinfo{address}{Oxford},
  \bibinfo{year}{1983}), \bibinfo{edition}{2nd} ed.

\bibitem[{\citenamefont{Breuer and Petruccione}(2002)}]{Bre02}
\bibinfo{author}{\bibfnamefont{H.~P.} \bibnamefont{Breuer}} \bibnamefont{and}
  \bibinfo{author}{\bibfnamefont{F.}~\bibnamefont{Petruccione}},
  \emph{\bibinfo{title}{The Theory of Open Quantum Systems}}
  (\bibinfo{publisher}{Oxford University Press}, \bibinfo{address}{Oxford},
  \bibinfo{year}{2002}).

\bibitem[{\citenamefont{Ubbelohde et~al.}(2012)\citenamefont{Ubbelohde, Fricke,
  Flindt, Hohls, and Haug}}]{Ubb121}
\bibinfo{author}{\bibfnamefont{N.}~\bibnamefont{Ubbelohde}},
  \bibinfo{author}{\bibfnamefont{C.}~\bibnamefont{Fricke}},
  \bibinfo{author}{\bibfnamefont{C.}~\bibnamefont{Flindt}},
  \bibinfo{author}{\bibfnamefont{F.}~\bibnamefont{Hohls}}, \bibnamefont{and}
  \bibinfo{author}{\bibfnamefont{R.~J.} \bibnamefont{Haug}},
  \bibinfo{journal}{Nature Communications} \textbf{\bibinfo{volume}{91}},
  \bibinfo{pages}{1} (\bibinfo{year}{2012}).

\end{thebibliography}

\end{document}